\documentclass[doublecol]{epl2}

\title{Stochastic resonance with weak monochromatic driving: gains above unity induced by high-frequency signals}

\author{Jes\'us Casado-Pascual\inst{1} \and David Cubero\inst{1} \and Jos\'e Pablo Baltan\'as\inst{2}}

\institute{                    
  \inst{1} F\'{\i}sica Te\'orica, Universidad de Sevilla - Apartado de Correos
1065, Sevilla 41080, Spain\\
  \inst{2} Departamento de F\'{\i}sica Aplicada II, Universidad de Sevilla - Avda.~Reina Mercedes, 2, Sevilla 41012, Spain
}

\abstract{
We study the effects of a high-frequency (HF) signal on the response of a noisy
bistable system to a low-frequency subthreshold sinusoidal signal. We show that, by conveniently choosing the ratio  of the amplitude of the HF signal to its frequency, 
stochastic resonance gains greater than unity can be measured at the low-frequency value. Thus, the addition of the HF signal can entail an improvement in the detection of weak monochromatic signals. The results are explained in terms of an effective model and illustrated by means of numerical simulations.
}

\begin{document}

\maketitle

The phenomenon of stochastic resonance (SR) has been studied with
growing interest during the last three decades, being found
to be of relevance in a great variety of phenomena in physics,
chemistry, and the life sciences~\cite{gamjun98}. Roughly speaking, SR
consists in the amplification of a weak, time-dependent signal of
interest by the concerted actions of noise and the nonlinearity of the
system. Several quantifiers have been used to characterise the SR
response of noisy systems in the presence of periodic signals. In
particular, the nonmonotonic behaviour of the output signal-to-noise
ratio (SNR) with the strength of the noise is a widely accepted
signature of SR. In addition, a dimensionless quantity known as the SR
gain is usually defined as the ratio of the output SNR over the input
SNR. The SNR measures the ``quality'' of the signal, in terms of the
ratio of its ``coherent'' (periodic) component over its ``incoherent''
(noisy) component. In turn, the SR gain compares the ``qualities'' of the
output and the input signals. In general, obtaining high output
SNRs and SR gains greater than unity would be desirable when using SR
as an amplification mechanism. Analog~\cite{loergin96,ginmak01} and
numerical~\cite{casdenk03,casgom_pre_03} simulations have shown that
noisy bistable systems driven by subthreshold multifrequency forces
can display SR gains greater than unity when the parameters of the
problem are properly chosen. Moreover, in~\cite{casgom03}, a
two-state model of SR has been used to explain these results
analytically. By contrast, to the best of our knowledge, there is no
evidence of SR gains greater than unity when a subthreshold monochromatic
signal drives an isolated bistable system. However, these unusual
large gains have been reported in the case of a suprathreshold
sinusoidal signal for an isolated bistable system~\cite{hanin00} and in
the case of a subthreshold sinusoidal signal for coupled
bistable systems~\cite{casgom06}.

As it has been already pointed out in the literature~\cite{hanin00},
an improvement of the response of a nonlinear system to a subthreshold
signal of interest embedded in noise can be achieved better by
lowering the threshold value rather than by increasing the noise
strength. However, in most cases of practical interest, threshold
lowering is usually a difficult task. Recently, it has been shown
that the effect of a strong high-frequency (HF) monochromatic force on the
overdamped dynamics of a Brownian particle in a bistable potential can
be described in terms of an effective potential whose characteristics
depend on the parameters of the HF field~\cite{casbal04}. In particular, the barrier height of this effective
potential is smaller than the original one. Thus, an HF
force provides a mechanism to achieve an effective threshold
lowering. As shown in~\cite{chigia05} for a {\em
square-wave} signal of interest, it is possible to take advantage of
this effective threshold lowering to improve the SR gain values
evaluated at the fundamental frequency of the square wave signal. More
precisely, in~\cite{chigia05}, SR gains greater than unity are
found for a wide range of values of the noise strength. However, as
mentioned in the same reference, this positive effect of the
HF force seems to be absent when the signal of interest is
{\em sinusoidal}. It is pertinent to point out that the effect of HF fields on nonlinear stochastic systems is not necessarily positive  
(see, for instance,~\cite{bormar05} for a ratchet model and~\cite{cubero06} for an excitable model). Nevertheless, in ratchet models positive effects have been also found~\cite{bormar06}.

In this work we show that, by using a strong HF signal, SR
gains greater than unity can be also observed in a bistable noisy
system when the signal of interest is {\em subthreshold} and {\em
sinusoidal}. In order to do that, let us consider a stochastic system characterised by a
single degree of freedom $x(t)$, whose dynamics (in dimensionless
units) is described by the stochastic differential equation
\begin{equation}
\dot{x}(t)=-U^{\prime}[x(t)]+F(t)+\xi(t)\,.
\label{eq:lang}
\end{equation}
Herein, $\xi(t)$ is a Gaussian white noise of zero mean and
autocorrelation function $\langle \xi(t)\xi(s)\rangle=2D\delta(t-s)$,
$D$ being the noise strength, $F(t)$ is the sinusoidal driving
$F(t)=A\cos(\Omega t)$, 
and $U^{\prime}(x)$ denotes the derivative with respect to $x$ of the
quartic potential $U(x)=x^4/4-x^2/2$.
Henceforth, we will restrict ourselves to subthreshold
driving signals. More precisely, we will assume that $A\le
A_{\mathrm{th}}=2/\sqrt{27}$, where $A_{\mathrm{th}}$ is the static
threshold value. In this case, the time dependent potential
$U(x)-A\,x\cos(\Omega t)$ possesses two local minima separated by a local maximum
for any instant of time $t$. Notice that the dynamical threshold value
(defined as the maximum value of $A$ such that no interwell
transitions are possible in the absence of noise) always exceeds this
adiabatic threshold $A_{\mathrm{th}}$.

Equation~(\ref{eq:lang}) describes the overdamped
motion of a particle in a symmetric double well potential driven by a
sinusoidal force and noise. Alternatively,
$x^{({\mathrm{in}})}(t)=F(t)+\xi(t)$ may be interpreted as an input signal
consisting of a noisy term $\xi(t)$ and a periodic term $F(t)$ (signal
of interest). This input signal is transformed into an output signal
$x^{({\mathrm{out}})}(t)=x(t)$ after being processed by the nonlinear
device characterised by $U(x)$. In this context, SR can be understood as a mechanism in
which noise plays a positive role in the optimal detection of some
features of the signal of interest $F(t)$ in the output signal $x(t)$.

Several quantifiers have been used to characterise the SR
phenomenon. Here, we will only consider those quantifiers involving
the power spectral density (PSD) of the input and output signals,
namely, the SNR of the output signal and the SR gain. According to the
generalisation of the Wiener-Khinchine theorem to a periodically
driven stochastic process, the PSD of the output signal $x(t)$ is
given by~\cite{jung93}
\begin{equation}
S^{(\mathrm{out})}(\omega)=\frac{2}{\pi}\int_{0}^{\infty} \mathrm{d}\tau\,
C(\tau)\cos(\omega\tau),
\label{eq:sout}
\end{equation}
with $\omega\ge 0$, where $C(\tau)=T^{-1}\int_{0}^{T} \mathrm{d}t\,\langle
x(t+\tau)x(t)\rangle_{\infty}$, $\langle\cdots\rangle_{\infty}$
representing the average over the realizations of the noise after a
relaxation transient stage. It can be proved that
$S^{(\mathrm{out})}(\omega)$ consists of a series of deltalike spikes
located at the odd harmonics of the fundamental frequency $\Omega$,
superimposed on a background PSD \cite{gamjun98}. This background PSD (also called
incoherent part of the PSD) is given by
\begin{equation}
S^{(\mathrm{out})}_{\mathrm{incoh}}(\omega)=\frac{2}{\pi}\int_{0}^{\infty}
\mathrm{d}\tau\, C_\mathrm{incoh}(\tau)\cos(\omega\tau), 
\label{eq:sincoh}
\end{equation}
where $C_\mathrm{incoh}(\tau)=C(\tau)-C_\mathrm{coh}(\tau)$, with
$C_\mathrm{coh}(\tau)=T^{-1}\int_{0}^{T} \mathrm{d}t\,\langle
x(t+\tau)\rangle_\infty\langle x(t)\rangle_{\infty}$ (see, for
instance,~\cite{casgom05}). Then, the output SNR is defined as \cite{gamjun98,casgom05}
\begin{equation}
R^\mathrm{(out)}=\lim_{\epsilon\rightarrow
0^+}\frac{\int_{\Omega-\epsilon}^{\Omega+\epsilon}\mathrm{d}\omega\,
S^{(\mathrm{out})}(\omega)}{S^{(\mathrm{out})}_\mathrm{incoh}(\Omega)}\,.
\label{eq:snrout}
\end{equation}

The input PSD $S^{(\mathrm{in})}(\omega)$, as well as
$S^{(\mathrm{in})}_{\mathrm{incoh}}(\omega)$ and $R^\mathrm{(in)}$,
can be obtained from the above expressions by replacing $x(t)$ by
$x^{(\mathrm{in})}(t)$. In particular, it is easy to prove that the input
SNR reads 
\begin{equation}
R^\mathrm{(in)}=\frac{\pi A^2}{4D}.
\label{eq:snrin}
\end{equation}
Finally, the SR gain is defined as
\begin{equation}
G=\frac{R^\mathrm{(out)}}{R^\mathrm{(in)}}\,.
\label{eq:gain}
\end{equation}
As mentioned before, obtaining SR gains greater than unity would be desirable. For this purpose, let us introduce a strong HF signal of the form:
\begin{equation}
Y(t)=N\Omega r \cos(N \Omega t+\varphi)\,,
\label{eq:cs}
\end{equation}
where the parameter $N$ is a positive integer, $r$ is the ratio of the
amplitude of $Y(t)$ to its frequency, and $\varphi$ is an arbitrary
initial phase. We are interested in situations in which the parameters
$N\Omega r$ and $N\Omega$ appearing in this monochromatic force are
much larger than the rest of the parameters in the problem. That is
why $Y(t)$ is called a strong HF signal. This situation
can be formally achieved by taking the limit $N\rightarrow\infty$,
with the ratio $r$ kept fixed. The HF signal $Y(t)$ must
be summed to the right hand side of eq.~(\ref{eq:lang}), so that now
$x(t)$ fullfils the Langevin equation 
\begin{equation}
\dot{x}(t)=-U^{\prime}[x(t)]+F(t)+Y(t)+\xi(t)\,.
\label{eq:langHF}
\end{equation}
As a consequence, the input signal in our previous discussion must be
replaced by $x^{({\mathrm{in}})}(t)=F(t)+Y(t)+\xi(t)$. It is easy to
prove that the input PSD in the presence of the HF signal
[$S^{(\mathrm{in})}(\omega;r)$ with $r\ne 0$] can be expressed in
terms of that obtained in its absence [$S^{(\mathrm{in})}(\omega;0)$]
as
\begin{equation}
S^{(\mathrm{in})}(\omega;r)=S^{(\mathrm{in})}(\omega;0)+
\frac{N^{2}r^2\Omega^2}{2}\delta(\omega-N\Omega)\,.
\label{eq:sx_input}
\end{equation}
It should be emphasised that we are interested in the evaluation of
the input and output SNRs at the frequency $\Omega$ [frequency of the
signal of interest $F(t)$], $Y(t)$ being just a tool introduced to
improve the response features of the system at that frequency. In
addition, since in the definition of $R^{(\mathrm{in})}$ only the PSD
around the frequency $\Omega$ plays a role, it is clear from
eq.~(\ref{eq:sx_input}) that the input SNR is given by
eq.~(\ref{eq:snrin}) even in the presence of the HF signal
$Y(t)$, as in this case $N\gg 1$.

For reasons that will be clarified later, it is convenient to define
the new stochastic process $z(t)=x(t)-r\sin(N\Omega t+\varphi)$. After replacing this definition in eq.~(\ref{eq:sout}),
it can be proved that the output PSD, $S^{(\mathrm{out})}(\omega)$,
 and the PSD of the process $z(t)$, $S_z(\omega)$, are related
 according to
\begin{equation}
S^{(\mathrm{out})}(\omega)=S_{z}(\omega)+
2r\left[\frac{r}{4}-\mathrm{Im}(e^{-i\varphi}Z_N)\right]
\delta(\omega-N\Omega)\,,
\label{eq:sx_sxhat}
\end{equation}
where $Z_N$ is the Fourier component of the process $z(t)$ at the
frequency $N\Omega$, i.e., $Z_{N}=T^{-1}\int_{0}^{T} \mathrm{d}t\,\langle
z(t)\rangle_{\infty}e^{-i N \Omega t}$. From eq.~(\ref{eq:sx_sxhat})
it is clear that, in order to evaluate quantities which only depend on
the behaviour of the output PSD around the frequency $\Omega$ [as for
instance $R^{(\mathrm{out})}$] both processes, $x(t)$ and $z(t)$, are
equivalent as long as $N\ne 1$.

The dynamics of the stochastic process $z(t)$ is described by
the equation
\begin{equation}
\dot{z}(t)=-U^{\prime}[z(t)+r\sin(N\Omega
t+\varphi)]+F(t)+\xi(t)\, .
\label{eq:lang3}
\end{equation}
As we have mentioned before, we are interested in the limit
$N\rightarrow \infty$, with $r$ kept fixed. From eq.~(\ref{eq:lang3}),
it follows that the time derivative of the process $z(t)$ is at most
of order $1$ as $N\rightarrow\infty$. In this sense, we will say that
this process is {\em slow}. In contrast, according to the definition of $z(t)$, 
it is clear that, in this limit, the process $x(t)$ is
{\em fast}, in the sense that it is a highly oscillating process
around $z(t)$. As the process $z(t)$ is {\em slow}, a large number of
oscillations of the function $r\sin(N\Omega t+\varphi)$ appearing in
$U^{\prime}[z(t)+r\sin(N\Omega t+\varphi)]$ takes place before a
significant change in $z(t)$ occurs. Therefore, for $N\gg 1$, $z(t)$
must be almost independent of the phase $\varphi$ and, consequently,
$z(t,\varphi)\simeq\bar{z}(t)\mathrel{\mathop:}=(2\pi)^{-1}
\int_{0}^{2\pi}\mathrm{d}\varphi^{\prime}\,z(t,\varphi^{\prime})$, where we have written
explicitly the dependence of $z(t)$ with respect to $\varphi$. Taking
into account this property and carrying out the phase average in
eq.~(\ref{eq:lang3}), one obtains
\begin{equation}
\dot{\bar{z}}(t)=-U_{\mathrm{eff}}^{\prime}[\bar{z}(t)]+F(t)+\xi(t)\, ,
\label{eq:eff}
\end{equation}
where we have introduced the effective potential
\begin{eqnarray}
U_{\mathrm{eff}}[\bar{z}]&\mathrel{\mathop:}=&
\frac{1}{2\pi}\int_0^{2\pi}\mathrm{d}\varphi\,
U[\bar{z}+r\sin(N\Omega t+\varphi)]\nonumber \\
&=&\frac{\bar{z}^4}{4}-c(r)\frac{\bar{z}^2}{2}+\frac{r^2}{4}\left[\frac{3 r^2}{8}-1\right]\, ,
\label{eq:eff_pot}
\end{eqnarray}
with $c(r)=1-3r^2/2$. From this result, it is clear that the stability
of this effective potential depends on the ratio $r$. More precisely,
if $r<\sqrt{2/3}$ the potential is bistable, whereas if
$r\ge\sqrt{2/3}$ is monostable. Thus, an increase in $r$ leads to a
decrease in the effective barrier height and, eventually, to its
disappearance. 

\begin{figure}
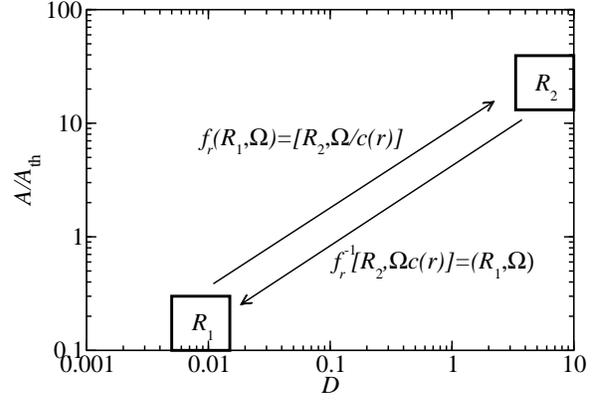

\onefigure[scale=0.30]{f1.eps}
\caption{Two regions, $R_1$ and $R_2$, of the parameter plane $D-A$ are transformed into each other by the mappings 
$f_r$ and $f_r^{-1}$. $R_1$ is the subthreshold region delimited by the four vertices
 $(D,A/A_{\mathrm{th}})=(0.005,0.1)$, $(0.015,0.1)$, $(0.015,0.3)$, and $(0.005,0.3)$, and $R_2$ is obtained 
from $R_1$ by applying $f_r$ with $r=0.800529$. This last parameter value has been chosen so
 that $R_2$ lies within a region wherein SR gains greater than
 unity have been observed in the case of a bistable system in the
 presence of a suprathreshold sinusoidal signal with frequency
 $\Omega=0.1$ (see fig.~2 in~\cite{hanin00}).}
\label{fig:scaling}
\end{figure}

Henceforth, we will restrict our study to the case $r<\sqrt{2/3}$,
which corresponds to $0<c(r)\le 1$ (bistable potential). In order to
reduce the effective Langevin equation [eq.~(\ref{eq:eff})] to the standard form [eq.~(\ref{eq:lang})], it is convenient to introduce the following time and space scale changes:
\begin{eqnarray}
\tilde{t}&=&t\,c(r)\, , \label{eq:t_resc}\\
\tilde{z}(\tilde{t})&=&\bar{z}
\left[\tilde{t}/c(r)\right]/\sqrt{c(r)}\, \label{eq:z_resc}.
\label{eq:xscale}
\end{eqnarray}
Then, it is easy to prove that the stochastic process $\tilde{z}(\tilde{t})$ fullfils the Langevin equation (\ref{eq:lang}) but for the rescaled values of the amplitude, noise strength, and frequency given by $A/c(r)^{3/2}$, $D/c(r)^{2}$, and $\Omega/ c(r)$, respectively. We will denote by $f_r$ the mapping which assigns to each set of parameter values $(A,D,\Omega)$ the corresponding rescaled values,
and by $f_r^{-1}$ its inverse mapping, i.e.,
\begin{eqnarray}
\label{mapping}
f_r(A,D,\Omega)&=&\left[A/c(r)^{3/2},D/c(r)^{2},\Omega/c(r)\right], \\
\label{inversemapping}
f_r^{-1}(A,D,\Omega)&=&\left[A c(r)^{3/2},D c(r)^{2},\Omega c(r)\right].
\end{eqnarray}
In fig.~\ref{fig:scaling} we depict how two regions of the parameter plane $D-A$, $R_1$ and $R_2$, are transformed into each other by the mappings $f_r$ and $f_r^{-1}$.
\begin{figure}
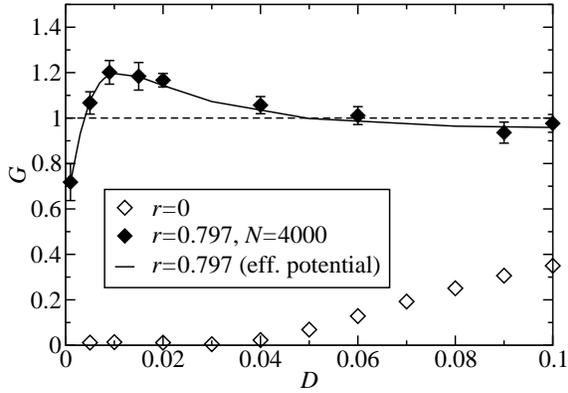

\onefigure[scale=0.30]{f2.eps}
\caption{Dependence on the noise strength $D$ of the SR gain $G$ in a bistable noisy system driven by a sinusoidal subthreshold signal, both in the absence (empty diamonds) and presence (full diamonds) of an HF signal. The parameter values of the sinusoidal subthreshold signal $F(t)$ and the HF signal $Y(t)$ are: $A=0.1$, $\Omega=0.005$, $r=0.797$, $N=4000$, and $\varphi=0$. The SR gain obtained from the numerical solution of the effective dynamics in eq.~(\ref{eq:eff}) is plotted with solid line. The horizontal dashed line marks the unity for the SR gain.}
\label{fig:sin}
\end{figure}

As mentioned before, for $N\ne 1$, the output SNR of the
processes $x(t)$ and $z(t)$ coincide exactly. Furthermore, for $N\gg
1$, the processes $z(t)$ and $\bar{z}(t)$ give rise to the same value
of this quantifier with a great deal of approximation. Finally, the
process $\bar{z}(t)$ is related to the process $\tilde{z}(\tilde{t})$
[which is the process $x(t)$ but in the absence of the
strong HF signal and for the above mentioned rescaled parameter values] by means of the transformations (\ref{eq:t_resc}) and (\ref{eq:z_resc}). Then, by using the
definition of the output SNR, one obtains that
\begin{equation}
R^{(\mathrm{out})}\left[(A,D,\Omega);r\right]\simeq c(r)R^{(\mathrm{out})}\left[f_r(A,D,\Omega);0\right]
\label{eq:rout_resc}
\end{equation}
for $N\gg 1$, as expected from the change of scale in eq.~(\ref{eq:t_resc}) and the fact that the SNR has dimensions of time$^{-1}$.
The above expression relates the output SNR in the presence of the
HF signal, $R^{(\mathrm{out})}[(A,D,\Omega);r]$, with that
corresponding to a new set of parameter values but in the absence of
the HF signal, $R^{(\mathrm{out})}\left[f_r(A,D,\Omega);0\right]$. 

Analogously, taking into account
the mapping in eq.~(\ref{mapping}), as well as
the fact that the input SNR at the frequency of interest $\Omega$ is
given by eq.~(\ref{eq:snrin}) even in the presence of the
HF signal [see the paragraph below
eq.~(\ref{eq:sx_input})], it is also clear that
\begin{eqnarray}
R^{(\mathrm{in})}[(A,D,\Omega);r]&=&R^{(\mathrm{in})}[(A,D,\Omega);0]\nonumber\\ &=&c(r)R^{(\mathrm{in})}\left[f_r(A,D,\Omega);0\right]\, .
\label{eq:rin_resc}
\end{eqnarray}
Finally, from eqs.~(\ref{eq:rout_resc}), (\ref{eq:rin_resc}), and the definition (\ref{eq:gain}), one obtains that
\begin{equation}
G[(A,D,\Omega);r]\simeq G\left[ f_r(A,D,\Omega);0\right]
\label{eq:gain_resc}
\end{equation}
for $N\gg 1$, this last equation being just a consequence of the dimensionless
character of the SR gain. Equation~(\ref{eq:gain_resc}) is one of the
main results of this paper. It shows that the value of the SR gain in
the presence of an HF signal is approximately equal to that
corresponding to a system in which that HF signal is
absent, but for a different set of parameter values. Obviously, the larger the value of $N$, the better the approximation. 

From eq.~(\ref{eq:gain_resc}) it is easy to prove that,  by applying an appropriate strong HF signal to a bistable noisy system driven by a subthreshold sinusoidal signal of interest, one can obtain SR gains greater than unity.
Indeed, as shown in~\cite{hanin00}, SR gains greater
than unity can be observed in a noisy bistable system driven by a
suprathreshold sinusoidal signal. For instance, in fig.~2 of~\cite{hanin00} it is shown that, for a fixed value  of the frequency of
this signal $\Omega=0.1$, there exists a region of the parameter plane $D-A$ wherein $G>1$. To fix ideas, let us assume that $G[(A_2,D_2,\Omega_2);0]>1$, with $A_2>A_{\mathrm{th}}$, and let us choose a value $r_0$ such that
$\sqrt{2\left[1-(A_{\mathrm{th}}/A_2)^{2/3} \right]/3}<r_0<\sqrt{2/3}$. If we define $(A_1,D_1,\Omega_1)=f_{r_0}^{-1}(A_2,D_2,\Omega_2)$, it is easy to check that $A_1<A_{\mathrm{th}}$. Furthemore, according to eq.~(\ref{eq:gain_resc}), $G[(A_1,D_1,\Omega_1);r_0]\simeq G[(A_2,D_2,\Omega_2);0]>1$, for $N\gg 1$. Thus, by applying an HF signal of the type given in eq.~(\ref{eq:cs}) (with $\Omega=\Omega_1$, $r=r_0$, $N\gg 1$, and an arbitrary value of $\varphi$)  to a noisy bistable system driven by a subthreshold sinusoidal signal of frequency $\Omega_1$ and amplitude $A_1$, one gets an SR gain greater than unity at a noise strength $D_1$. This is exactly what we wanted to prove. 

As an example of the above reasoning, let us consider fig.~\ref{fig:scaling}. There, $R_2$ lies within a region of the parameter plane $D-A$ wherein SR gains greater than
 unity have been observed in the case of a bistable system in the
 presence of a suprathreshold sinusoidal signal with frequency
 $\Omega=0.1$ (see fig.~2 in~\cite{hanin00}). Therefore,  according to our reasoning, it is expected to observe SR gains greater than unity within the subthreshold region $R_1$ and for the frequency value $\Omega=0.003873$, whenever one applies an HF signal of the type given in eq.~(\ref{eq:cs}), with  $r=0.800529$, $N\gg 1$, and an arbitrary value of $\varphi$.

In order to illustrate the feasibility of the above argument, in fig.~\ref{fig:sin} we depict the dependence on the noise strength $D$ of the SR gain in a bistable noisy system driven by a sinusoidal subthreshold signal, both in the absence (empty diamonds) and presence (full diamonds) of an HF signal. The SR gain has been evaluated numerically following the method described in~\cite{casdenk03}. The parameter values of the sinusoidal subthreshold signal are $A=0.1$ and $\Omega=0.005$. As expected, in the absence of the HF signal, the SR gain is lower than unity for all the noise strength values. Nevertheless, when an HF signal of the type given in eq.~(\ref{eq:cs}) is applied, with $\Omega=0.005$, $r=0.797$, $N=4000$, and $\varphi=0$, there is a
range of noise strength values for which SR gains greater than unity are observed, the maximum value of $G$ being located around $D_{\mathrm{max}}= 0.01$. Since our argument is premised on the equivalence of the exact and the effective dynamics [eqs.~(\ref{eq:langHF}) and (\ref{eq:eff}), respectively] for evaluating the SR quantifiers in the asymptotic limit $N\to \infty$, it is important to check the validity of this approximation for $N=4000$. In fig.~\ref{fig:sin}, we have also plotted with solid line the SR gain obtained from the numerical solution of the effective dynamics in eq.~(\ref{eq:eff}). A glance at that figure reveals that the agreement between both dynamics is excellent for $N=4000$.

It is worth noting that the rescaled parameter values of the monochromatic signal in fig.~\ref{fig:sin} [i.e., $A/c(r)^{3/2}\approx 25.347 A_{\mathrm{th}}$ and $\Omega/c(r)\approx 0.106$]  are almost equal to the amplitude and frequency of the suprathreshold sinusoidal signal of the dot-dashed line in fig.~5 of~\cite{hanin00}. Thus, according to eq.~(\ref{eq:gain_resc}), the above mentioned solid and  dot-dashed lines must be essentially the same, except for a rescaling of the $D$-axis by a factor $1/c(r)^{2}\approx 449.122$. So, for instance, the location of the maximum of the SR gain in the dot-dashed line is approximately equal to $D_{\mathrm{max}}/ c(r)^{2}\approx 4.491$, and the maximum value of the SR gain is approximately $1.2$ in both curves.

In conclusion, in this work we have shown that when a noisy input containing a subthreshold
monochromatic signal of interest is processed by a bistable system in the
presence of an HF signal, SR gains greater than unity can be
observed. In order to explain this result, we have proved that the original bistable
system in the presence of the HF signal can be approximated by a new
bistable system obeying the same stochastic dynamics as the original one but in the absence
of the HF signal. In this new system, the amplitude and
frequency of the signal of interest, as well as the noise strength, appear
rescaled by coefficients depending on the ratio $r$ of the amplitude of the
HF signal to its frequency. The main result of the paper
establishes that the SR gains of both systems are approximately equal. The
importance of this result relies on the fact that, although the
original parameter values lie within a subthreshold region, the rescaled parameter values obtained from a convenient choice of $r$ can belong to a suprathreshold region wherein SR gains greater than unity have been observed. As the SR gains in both
systems approximately coincide, it is apparent that the addition of the HF signal can give rise to SR gain values greater than unity. Finally, the theoretical results have been illustrated in a
particular case by means of numerical simulations. The authors are confident that the results reported in this paper can be useful for the optimal detection of weak monochromatic signals in the great variety of systems modeled by noisy bistable potentials.

\acknowledgments
The research was supported by the Direcci\'on General de Ense\~nanza
Superior of Spain (Grant No. FIS2005-02884) and the Junta de Andaluc\'{\i}a (J. C.-P. and D. C.), the Juan de la Cierva program of the Ministerio de Ciencia y Tecnolog\'{\i}a (D. C.), and the Consejer\'{\i}a de Educaci\'on y Ciencia de la Junta de Andaluc\'{\i}a (FQM-276) (J. P. B).

\end{document}